# The valence state of iron in the Sr$_2$Fe(Mo,W,Ta)O$_{6.0}$ double-perovskite system: An Fe *K*-edge and *L*$_{2,3}$-edge XANES study

M. Karppinen,[1]* H. Yamauchi,[1] Y. Yasukawa,[1] J. Lindén,[2] T. S. Chan,[3] R. S. Liu,[3] and J. M. Chen[4]

[1]*Materials and Structures Laboratory, Tokyo Institute of Technology, Yokohama 226-8503, Japan*
[2]*Physics Department, Åbo Akademi, FIN-20500 Turku, Finland*
[3]*Department of Chemistry, National Taiwan University, Taipei, Taiwan, Republic of China*
[4]*National Synchrotron Radiation Research Center, Hsinchu, Taiwan, Republic of China*

*Corresponding author (e-mail: karppinen@msl.titech.ac.jp)

Here we employ both Fe *K* and *L*$_{2,3}$ edge x-ray absorption near-edge structure (XANES) spectroscopy techniques to clarify that iron in the *B*-site ordered double-perovskite halfmetal, Sr$_2$FeMoO$_{6.0}$, possesses a mixed-valence state, Fe$^{II/III}$, and accordingly molybdenum a mixed Mo$^{V/VI}$ valence state. A reliable interpretation of the spectral features has been made possible by using a series of samples of the Sr$_2$Fe(Mo,W/Ta)O$_{6.0}$ system. Replacing Mo$^{V/VI}$ gradually with W$^{VI}$ causes increasing amount of Fe to adopt the Fe$^{II}$ state, whereas Ta$^V$ substitution shifts the valence of iron towards Fe$^{III}$. As the valence of Fe increases from II to III in the Sr$_2$Fe(Mo,W/Ta)O$_{6.0}$ system, the absorption energy at the Fe *K*-edge gradually shifts towards the higher energy side. Similarly, in the *L*$_{2,3}$-edge XANES spectra intermediate spectral features are revealed for the Sr$_2$FeMoO$_{6.0}$ sample in comparison with those for samples heavily substituted with either W or Ta.

**Introduction**

The *B*-site ordered double perovskites, $A_2BB'O_{6-w}$, are derived from the simple perovskite compounds, $ABO_3$, upon co-occupation of the octahedral cation site with two different metal species of different charges. Such category of compounds has been re-highlighted since room-temperature halfmetallicity and tunneling-type magnetoresistance behavior were revealed for $Sr_2FeMoO_{6-w}$,[1] $Sr_2FeReO_{6-w}$,[2] and related $A_2Fe(Mo,Re)O_{6-w}$ (*A* = Ca, Sr, Ba) compounds. Long before the present rush in research of these compounds they were known as conductive room-temperature ferrimagnets.[3] In an ideal double perovskite the two *B*-site cations, *B* and *B'*, are perfectly ordered such that each $BO_6$ octahedron is surrounded by six corner-sharing $B'O_6$ octahedra, and *vice versa*. For the different pairs of *B* and *B'*, the larger the charge difference is, the higher is the degree of order at the *B* site.[4]

Actual valence states of the *B*-site cations in $A_2Fe(Mo,Re)O_{6-w}$ were discussed for the first time in 1970's.[5] Renewed debates on this topic have been actively going on for a couple of years.[6,7] The $Sr_2FeMoO_{6-w}$ phase forms only under strongly reduced oxygen partial pressures, underlining the fact that either one (or both) of the *B*-site cation constituents possesses a relatively low valence value, *i.e.*, lower than III for Fe and/or VI for Mo.[8] Initially a picture was assumed based on high-spin $Fe^{III}$ with localized $3d^5$ electrons ($t_{2g}^3 e_g^2$; $S = 5/2$) and $Mo^V$ with an itinerant $4d^1$ electron ($t_{2g}^1$; $S = 1/2$). However, we recently proposed that iron is in a mixed-valence or valence-fluctuating state expressed as $Fe^{II/III}$, based on the $^{57}Fe$ Mössbauer spectra revealing intermediate hyperfine parameters for Fe in $Sr_2FeMoO_{6.0}$ as compared with those typically obtained for high-spin $Fe^{III}$ and high-spin $Fe^{II}$.[9] Note that molybdenum should be at a mixed $Mo^{V/VI}$ valence state accordingly, as the itinerant *d* electron of Mo transfers part of its charge and spin density to Fe. The mixed valence interpretation has been accepted in other $^{57}Fe$ Mössbauer studies,[10,11] though divergent interpretations have been reported as well.[12] The reported $L_{2,3}$-edge x-ray absorption near-edge structure (XANES) spectra for $Sr_2FeMoO_{6-w}$ have been ascribed to both $Fe^{III}$ [13,14] and $Fe^{II/III}$ [15]. For Mo, an NMR study suggested a mixed-valence state, $Mo^{V/VI}$.[16]

Here, using a series of samples of the $Sr_2Fe(Mo,W/Ta)O_{6-w}$ system we are able to show that the Fe XANES spectra not only at the $L_{2,3}$ edge but also at the *K* edge may be interpreted on the basis of actual II/III mixed-valence state for iron in $Sr_2FeMoO_{6.0}$. The end-members of the sample series, $Sr_2FeWO_6$[17] and $Sr_2FeTaO_6$[18], provide us with optimal references for iron of pure divalent and pure trivalent state, respectively, in an oxide environment as akin as possible to that in $Sr_2FeMoO_{6.0}$. Replacing $Mo^{V/VI}$ with increasing amount of $W^{VI}$ thus makes the valence of Fe gradually approach the II state, whereas that with increasing amount of $Ta^V$ causes continuous shift of the Fe valence towards III. The very same samples used in the present study previously had undergone thorough characterizations by means of $^{57}$Fe Mössbauer spectroscopy for the hyperfine parameters of different Fe species present in the samples and transport measurements under different magnetic fields for the magnetoresistance characteristics, the results of which were reported elsewhere.[19,20] For the samples heavily substituted with either W or Ta enhanced low-temperature MR values were achieved. For this reason the $Sr_2Fe(Mo,W)O_{6-w}$ system was already highlighted in earlier studies.[21,22]

**Experimental Section**

**Sample Preparation.** The $Sr_2Fe(Mo,W/Ta)O_{6-w}$ samples used in this study were synthesized by means of an oxygen-getter-controlled low-$O_2$-pressure encapsulation technique.[8] As starting materials, stoichiometric mixtures of high-purity $SrCO_3$, $Fe_2O_3$, $MoO_3$ and $WO_3$/$Ta_2O_5$ powders were used. Calcinations were carried out for the thoroughly mixed powders in an Ar atmosphere at 900 °C for 15 hours. The calcined powders were pelletized and sintered in evacuated fused-quartz ampoules containing sample pellets together with Fe grains (99.9 % up, under 10 mesh) as a getter of excess oxygen. The empty space inside the ampoule was filled with a fused-quartz rod. The synthesis was carried out at 1150 °C for 50 hours. At 1150 °C the oxygen partial pressure equilibrates in the presence of the Fe/FeO redox couple at $\sim 2.6 \times 10^{-13}$ atm.[23]

**Characterization.** The synthesized samples were checked for phase-purity and lattice parameters by x-ray diffraction (XRD; MAC Science M18XHF[22]; Cu$K_\alpha$ radiation). For the oxygen-content determination, an analysis method[8] based on coulometric titration of $Fe^{II}$ and/or $Mo^V$ species formed upon acidic dissolution of the sample was applied. The samples were also characterized for the dc magnetization (from 5 K to 400 K, and –5 T to 5 T) using a superconductivity-quantum-interface-device magnetometer (SQUID; Quantum Design: MPMSR-5S). From the measured magnetization data the magnitude of saturation magnetization ($M_S$) was determined as the magnetization value *per* formula unit at 5 T and 5 K.

**X-ray Absorption Measurements.** The Fe $K$- and $L_{2,3}$-edge XANES measurements were performed at National Synchrotron Radiation Research Center (NSRRC) in Hsinchu, Taiwan with an electron beam energy of 1.5 GeV and a maximum stored current of 240 mA.[24] For the XANES measurement the sample was ground to pass through a 400-mesh sieve to fulfil the requirement that the size of the particles is smaller than the absorption length in the material, *i.e.* $\mu d$ < 1, where $d$ is the particle size and $\mu$ is the total absorption coefficient. The resultant fine powder was rubbed homogeneously onto Scotch tape. Then the thickness of the sample was carefully adjusted by folding the Scotch tape several times to achieve $\Delta\mu x \approx 1$, where $\Delta\mu x$ is the edge step. The measurements at the Fe $K$ edge were performed in transmission mode at the Wiggler beamline BL-17C with a double-crystal Si (111) monochromator. Gas-ionization chambers filled with gas mixtures of $N_2$-He and $N_2$-Ar were used as detectors to measure (with a scan step of ~0.4 eV in the XANES region), respectively, the incident ($I_0$) and transmitted ($I$) photon intensities. As a reference for energy calibration, the spectrum of Fe metal foil was simultaneously monitored. The higher x-ray harmonics were minimized by detuning the double-crystal Si(100) monochromator to 80 % of the maximum. The Fe $L_{2,3}$-edge spectra were recorded by measuring the sample drain current in an ultrahigh vacuum chamber ($10^{-9}$ torr) at the 6-m high-energy spherical grating monochromator (HSGM) beamline. The incident photon flux ($I_0$) was monitored simultaneously by using a Ni mesh located after the exit slit of the monochromatic beam. The absorption measurements were normalized to $I_0$. The reproducibility of the adsorption spectra of the same sample in different experimental runs was found to be extremely good. All the measurements were performed at room temperature.

## Results and Discussion

The $Sr_2Fe(Mo,W/Ta)O_{6-w}$ samples were found to be phase pure for the whole Ta-for-Mo substitution range and for W-for-Mo substitution up to ~70 % substitution level. For the higher W contents traces of the non-magnetic Scheelite-type $Sr(Mo,W)O_4$ phase were detected in the x-ray diffraction patterns. On the other hand, for the Ta-substituted samples the superlattice peaks due to doubling of the lattice parameter of the simple perovskite cell were clearly distinguished up to ~80 % substitution level only, indicating that $Sr_2FeTaO_{6-w}$ rather possesses a simple-perovskite structure with random occupation of the $B$-lattice site by Fe and Ta atoms. For the non-substituted $Sr_2FeMoO_{6-w}$ sample the measured saturation magnetization, $M_S$, was as high as 3.5 $\mu_B$ reflecting the high degree of order among the Fe and Mo atoms. (The reduction in $M_S$ as compared with the maximum value of $4\mu_B$ is commonly attributed to the decreased degree of order at the *B*-cation site.) Parallel with the high $M_S$ value, our $^{57}$Fe Mössbauer spectroscopy investigation carried out for the same Sr$_2$FeMoO$_{6-w}$ sample had revealed that the fraction of Fe atoms occupying the "wrong" Mo site is as low as 3 %.[20] Here it should also be noted that Mössbauer data have clearly shown that the misplaced or so-called antisite Fe atoms in the Sr$_2$Fe(Mo,W/Ta)O$_{6-w}$ system are trivalent.[9,11,20] Whereas the Ta substitution was found to decrease the degree of *B*-site order, substituting Mo with W enhanced the ordering. This trend is rather what one expects, since for the *B*-cation pair of Fe$^{II}$ and W$^{VI}$ the charge difference is 4 while it is only 2 for Fe$^{III}$ and Ta$^{V}$. For the both sample series, Sr$_2$Fe(Mo,W)O$_{6-w}$ and Sr$_2$Fe(Mo,Ta)O$_{6-w}$ the unit volume of the lattice as determined from the XRD patterns showed monotonous evolution as the W/Ta substitution proceeded. The volume *versus* substitution level plots were given in Ref. 20 and are therefore not repeated here. The small deviations seen in these plots from the completely linear behavior are well explained by the changes in the degree of *B*-site cation order. (For *B*-site ordered double perovskites the lattice dimension(s) are typically found to expand with decreasing degree of order.[4])

For all the W-substituted samples and also for the Ta-substituted samples up to 40 % substitution level, the oxygen content was precisely determined by means of wet-chemical analysis. For the heavily Ta-substituted samples the analysis was not possible due to poor solubility of the sample material. For all the samples investigated, the analysis yielded a stoichiometric value of 6.00 for the oxygen content *per* formula unit within the error limits of ±0.03. Consistently with this result our previous Mössbauer data had not revealed any indication of five-fold coordinated Fe species for any other sample except for that substituted with 100 % Ta.[20] For Sr$_2$FeTaO$_{6-w}$, 4 % of Fe atoms were found to possess the (five-coordinated) divalent state,[20] thus suggesting oxygen deficiency of $w \approx 0.02$ (that is not significant either).

In Figure 1 we show the $L_3$ portion of the Fe $L_{2,3}$-edge XANES spectra for a series of Sr$_2$Fe(Mo,W/Ta)O$_{6.0}$ samples. The main spectral features of the $L_{2,3}$ edge of Fe originate from dipole transitions from the core Fe $2p$ level to the empty Fe $3d$ states.[25,26] The spectra are separated into two regions due to core-hole spin-orbit interaction: Fe $2p_{3/2}$ ($L_3$ edge; 705 ~ 715 eV) and Fe $2p_{1/2}$ ($L_2$ edge; 715 ~ 730 eV). Transitions $2p \rightarrow 4s$ are also allowed but much weaker contributing only to the smooth background at higher energies. Both the edges, $L_3$ and $L_2$, are further divided into two peaks. The splitting and intensity ratio between the two peaks is determined by the interplay of crystal-field effects and electronic interactions. The $L_3$ absorption edge of $Fe^{II}$ species in an octahedral crystal field typically exhibits a main peak at a lower energy (~707 eV), followed by a weaker peak or a shoulder at a higher energy (~709 eV).[25,26] The order of the peaks is reversed for $Fe^{III}$ species. This is what is precisely seen for the $Sr_2Fe(Mo,W/Ta)O_{6.0}$ sample series: the lower-energy peak is stronger than the higher-energy peak for heavily $W^{VI}$-substituted samples and weaker for the heavily $Ta^V$-substituted ones. The $Sr_2FeMoO_{6.0}$ sample possesses intermediate spectral features as compared with those for the strongly W- and Ta-substituted samples, as a manifestation of the $Fe^{II/III}$ mixed-valence state in it. For a more quantitative illustration, we approximate the intensities, $I_{707}$ and $I_{709}$, of the two $L_3$-edge peaks by the peak heights and plot the intensity ratio, $I_{709}/I_{707}$, against the W/Ta substitution level in Figure 1(b).

In Figure 2(a), the Fe $K$-edge absorption spectra are shown for selected $Sr_2Fe(Mo,W/Ta)O_{6.0}$ samples together with those for three simple iron oxides as references: $Fe^{II}O$ (wüstite of the rock-salt structure with an octahedral site for $Fe^{II}$), $Fe^{II/III}_3O_4$ (magnetite of the inverse-spinel structure with an octahedral site for $Fe^{II}$ and tetrahedral and octahedral sites for $Fe^{III}$) and α-$Fe^{III}_2O_3$ (hematite of the corundum structure with an octahedral site for $Fe^{III}$). For the reference oxides, it is concluded that the main peak that is due to transitions $1s \rightarrow 4p$ clearly shifts to the higher energy with an increasing valence state of iron. A small prepeak is seen at the low-energy side of the main peak for all the three oxides. This is probably due to transitions $1s \rightarrow 3d$, though the exact origin of the fine structure seen at the Fe $K$ edge is still being debated.[26] The spectra of the $Sr_2Fe(W_{0.9}Mo_{0.1})O_{6.0}$, $Sr_2Fe(W_{0.2}Mo_{0.8})O_{6.0}$, $Sr_2Fe(Mo_{0.8}Ta_{0.2})O_{6.0}$ and $Sr_2FeTaO_{6.0}$ samples also shown in Figure 2(a) are located roughly between those for $Fe^{II}O$ and α-$Fe^{III}_2O_3$, obeying the trend seen for the reference oxides, *i.e.*, with increasing valence state of iron the main absorption edge shifts to the higher energy. The details of the spectral features of the $Sr_2Fe(Mo,W/Ta)O_{6.0}$ samples somewhat differ from those of the simple iron oxides, whereas within the $Sr_2Fe(Mo,W/Ta)O_{6.0}$ double perovskite series, the spectral features evolve smoothly. This underlines the fact that the sample series itself provides us with the best reference system. The very edge area of the spectra from 7115 to 7130 eV is shown in Figure 2(b) for the whole series of $Sr_2Fe(Mo,W/Ta)O_{6.0}$ samples. Clearly, the absorption edge monotonously shifts to the higher energy first with decreasing amount of $W^{VI}$ that replaces $Mo^{V/VI}$ in $Sr_2Fe(Mo,W)O_{6.0}$ and then with increasing amount of $Ta^V$ to replace $Mo^{V/VI}$ in $Sr_2Fe(Mo,Ta)O_{6.0}$, *i.e.*, with the expected valence of Fe increasing from II to III. Thus, for the Mo-rich samples, absorption energy values that are intermediate between those for strongly $W^{VI}$- or $Ta^V$-substituted samples are seen, which manifests the intermediate valence state of iron in these samples. From Figure 2(b), even though the overall trend is clear within the whole $Sr_2Fe(Mo,W/Ta)O_{6.0}$ system, the shift of the $K$-edge absorption energy is larger for the $W^{VI}$-substituted samples as compared with those substituted with $Ta^V$. The small oxygen-deficiency of $w \approx 0.02$ concluded for the 100 % Ta-substituted sample from the Mössbauer data provides us with partial (though not complete) explanation. The observed behavior at the $K$ edge might thus indicate that the actual valence of Fe in $Sr_2FeMoO_{6.0}$ is not precisely 2.5 but a little higher.

In conclusion, we employed Fe XANES spectroscopy at both $K$ and $L_{2,3}$ edges to show that iron in the halfmetallic $Sr_2FeMoO_{6.0}$ magnetoresistor possesses an $Fe^{II/III}$ mixed-valence state. The key for the reliable interpretation of the spectral features was to use a full series of double-perovskite samples, $Sr_2Fe(Mo,W/Ta)O_{6.0}$, as a reference system where the valence of Fe gradually varies from II to III. At both the edges, $K$ and $L_{2,3}$, intermediate spectral features were revealed for Mo-rich samples as compared with those for the strongly W- and Ta-substituted samples. The mixed-valence state of iron as thus confirmed for $Sr_2FeMoO_{6.0}$ is in line not only with the Mössbauer data[9,19,20] but also with the result of band-structure calculation[1], suggesting a strong mixing of the itinerant $d$ electron from nominally pentavalent Mo and the minority spin $t_{2g}$ band of nominally trivalent Fe. It also agrees with the neutron diffraction data showing reduced magnetic moment values at both the Fe and Mo sites from the values expected for high-spin $Fe^{III}$ ($t_{2g}^3 e_g^2$; $S = 5/2$) and $Mo^V$ ($t_{2g}^1$; $S = 1/2$).[10] Finally we like to emphasize that perovskite-derived mixed-valent Fe oxides are gaining considerable research interest owing to the phenomena such as charge separation and ordering in $(La,Sr)FeO_3$[27] and in $REBaFe_2O_5$[28] ($RE$ is rare earth element) and large magnetoresistance in $SrFeO_{2.95}$[29] that are intimately related to the valence of iron. Thus the results of the present work are expected to be of wider importance in understanding the (mixed) valence state of iron in a larger group of perovskite oxides, as probed by various experimental techniques.

**Acknowledgments.** The present work has been supported by a Grant-in-Aid for Scientific Research (contract No. 11305002) from the Ministry of Education, Science and Culture of Japan, and also through the JSPS Research Fellowship Program for Young Scientists (No. 14006635; Y.Y.). T. Yamamoto is thanked for his contribution in sample preparation.

**Figure Captions**

**Fig. 1.** (a) Fe $L_3$-edge XANES spectra for a series of $Sr_2Fe(Mo,W/Ta)O_{6.0}$ samples in the energy range of 704 - 714 eV, and (b) the intensity ratio, $I_{709}/I_{707}$, of the two peaks at ~707 and ~709 eV (approximated by the peak heights) as plotted against the W/Ta substitution level.

**Fig. 2.** Fe $K$-edge XANES spectra (a) for the selected $Sr_2Fe(Mo,W/Ta)O_{6.0}$ samples and the reference oxides, $Fe^{II}O$, $Fe^{II/III}_3O_4$ and $Fe^{III}_2O_3$ in the energy range of 7100 - 7150 eV, and (b) for all the $Sr_2Fe(Mo,W/Ta)O_{6.0}$ samples in the energy range of 7115 - 7130 eV.

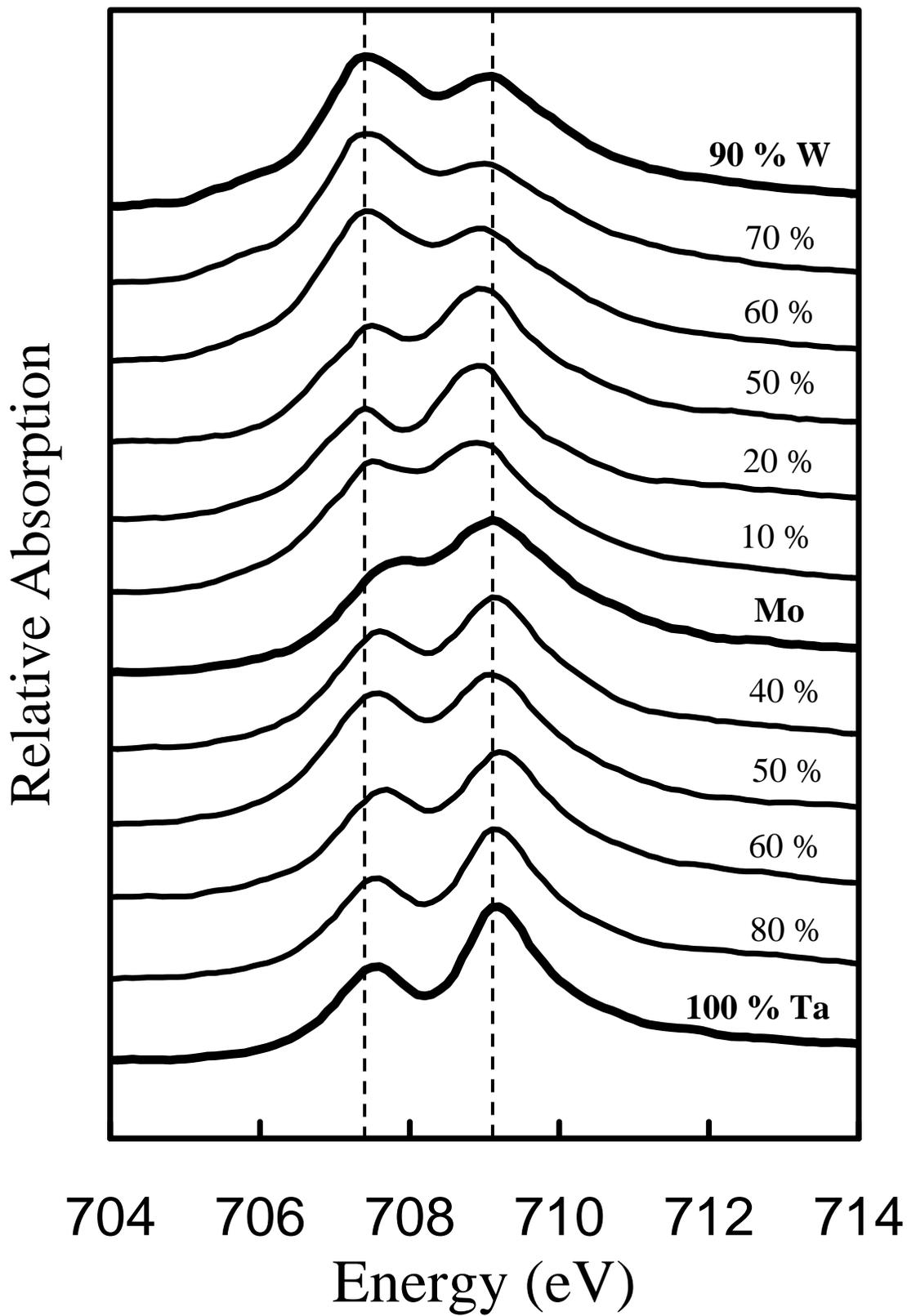

Karppinen *et al*: **Fig. 1(a).**

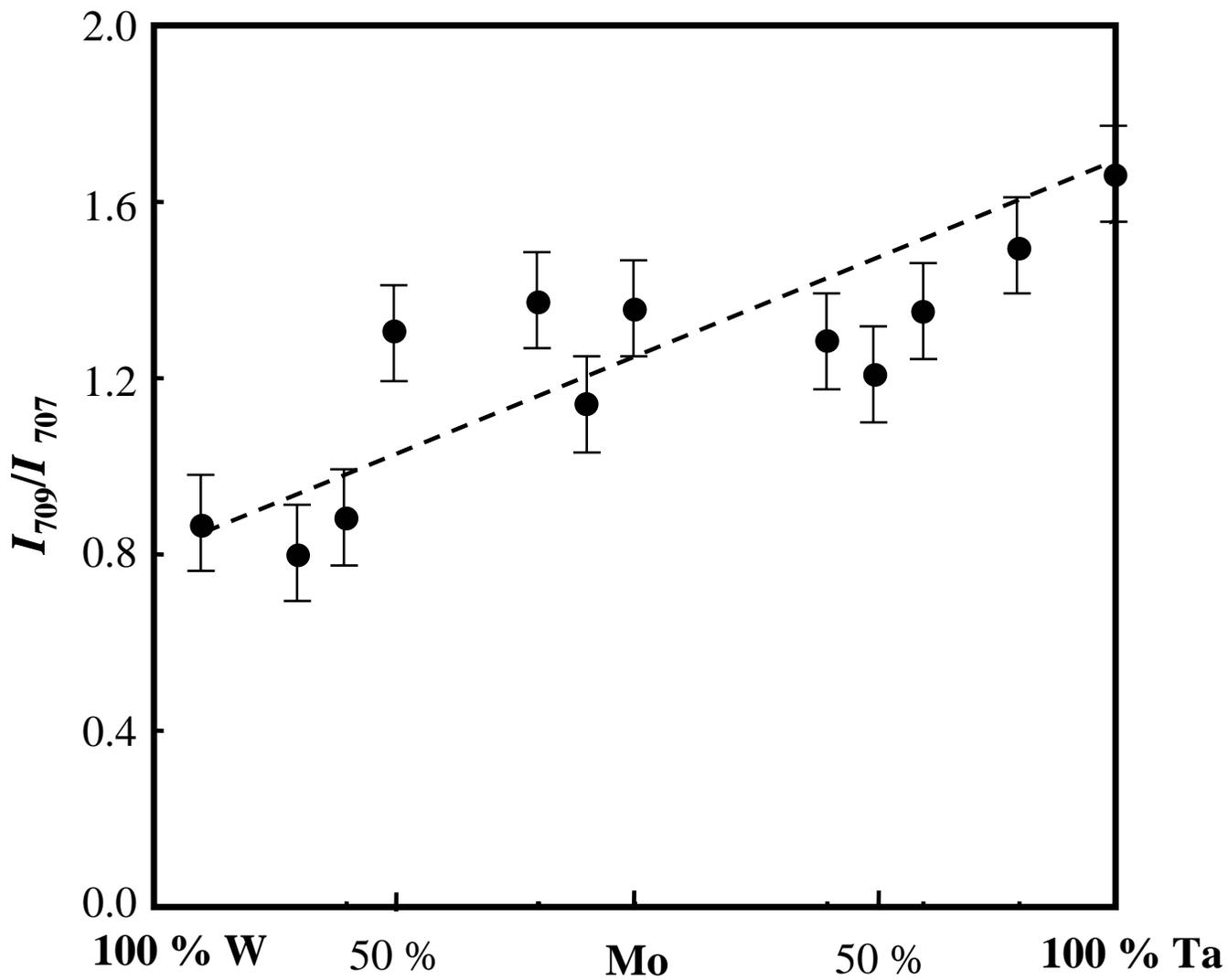

Karppinen *et al*: **Fig. 1(b).**

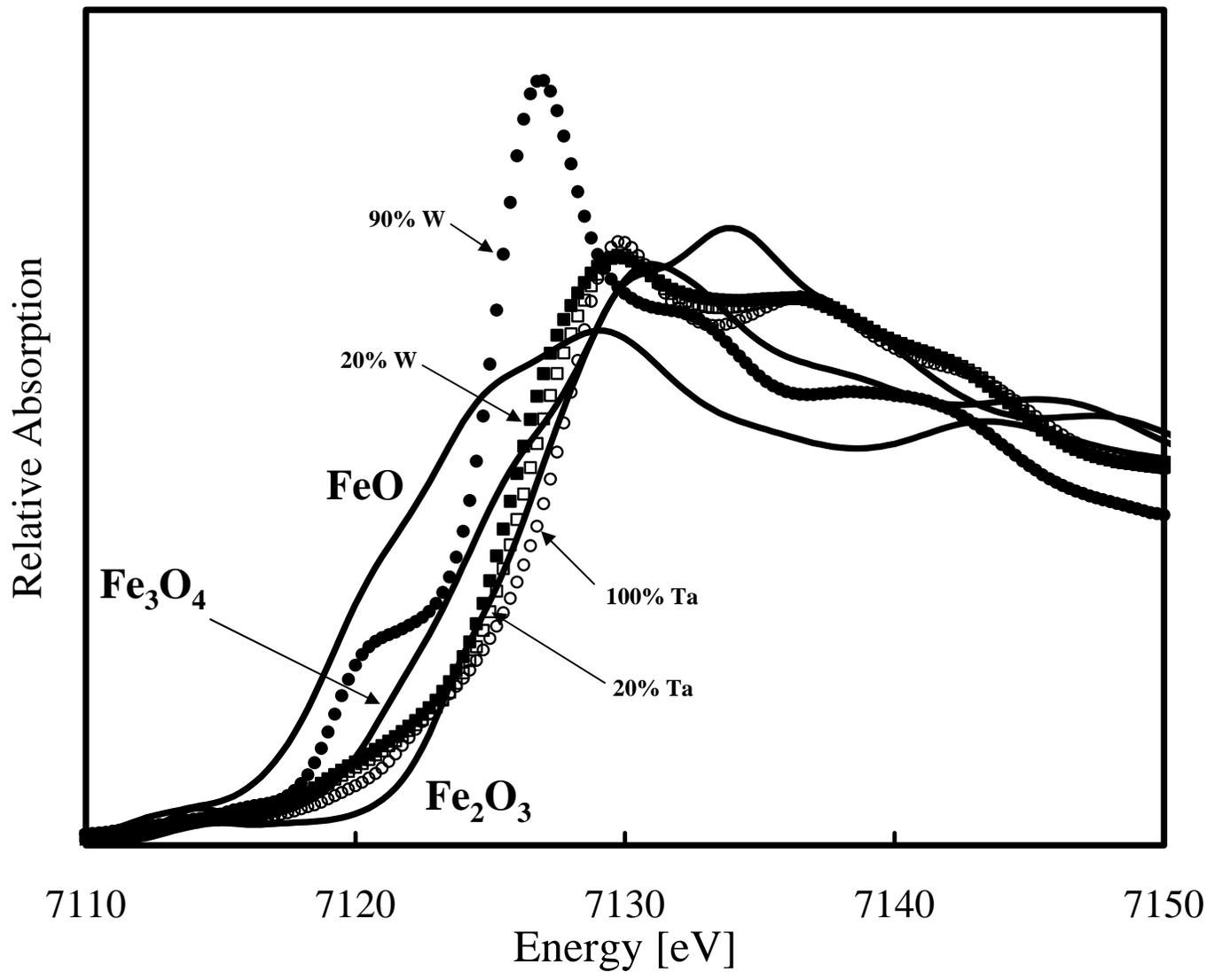

Karppinen *et al*: **Fig. 2(a).**

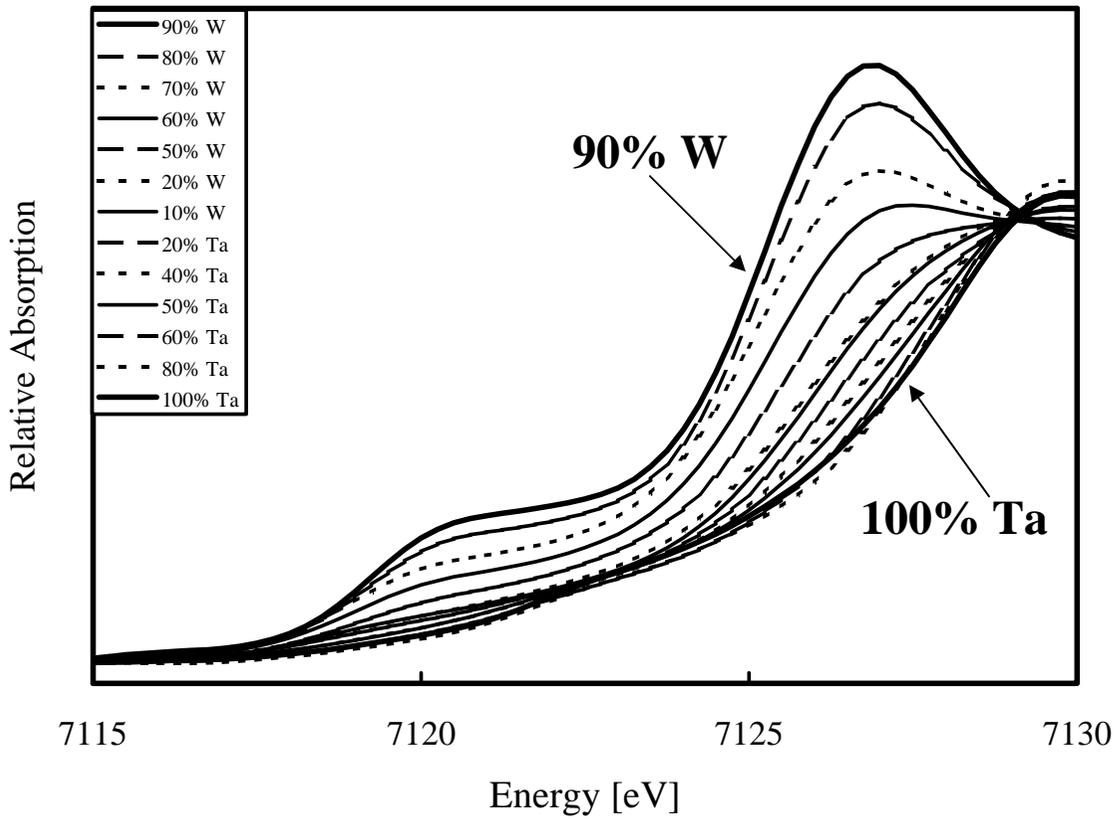

Karppinen *et al*: **Fig. 2(b).**